\documentclass[hyper]{prop2015}
\usepackage[english]{babel}
\usepackage[all,knot]{xy}
\usepackage{mathrsfs}
\usepackage{bbm}
\usepackage{framed}
\newcommand{\CA}{\mathcal{A}}    			

\newcommand{\CC}{\mathcal{C}}
\newcommand{\CCC}{\mathscr{C}}

\newcommand{\CD}{\mathcal{D}}

\newcommand{\CF}{\mathcal{F}}

\newcommand{\CCG}{\mathscr{G}}
\newcommand{\CH}{\mathcal{H}}

\newcommand{\CI}{\mathcal{I}}

\newcommand{\CL}{\mathcal{L}}
\newcommand{\CM}{\mathcal{M}}

\newcommand{\CQ}{\mathcal{Q}}

\newcommand{\CV}{\mathcal{V}}

\newcommand{\CX}{\mathcal{X}}

\newcommand{\CY}{\mathcal{Y}}

\newcommand{\CZ}{\mathcal{Z}}

\newcommand{\CE}{\mathcal{E}}
\newcommand{\CCE}{\mathscr{E}}

\newcommand{\pr}{\mathsf{pr}}
\newcommand{\dd}{\mathrm{d}}   

\category{Proceedings}
\subtitle{\href{http://www.maths.dur.ac.uk/lms/109/index.html}{LMS/EPSRC Durham Symposium on Higher Structures in M-Theory}}

\title{Pre-NQ Manifolds and Correspondence Spaces: the Nilmanifold Example}

\author[A. Deser]{Andreas Deser\inst{a,}\footnote{Corresponding author e-mail:~\href{mailto:andreas3deser@gmail.com}{\textsf{andreas3deser@gmail.com}}}}
\address[1]{Faculty of Mathematics and Physics,
  Charles University,
  Sokolovsk{\'a} 83, 186 75 Praha 8, 
  Czech Republic}
  
  \begin{acknowledgements}
The author is grateful to Jim Stasheff for discussion. The work leading to this proceedings has been partially supported by the COST (European Cooperation in Science and Technology) Action MP1405 QSPACE. Furthermore, this research was supported by OP RDE project No.~CZ.02.2.69/0.0/0.0/16\_027/0008495, International Mobility of Researchers at Charles University.
\end{acknowledgements}

\shortauthors{A. Deser}
\begin{abstract}
  Courant algebroids correspond to degree-2 symplectic differential graded manifolds or NQ-manifolds for short. We review how a similar construction shows that locally the gauge structure of Double Field Theory corresponds to degree-2 symplectic pre-NQ manifolds. To illustrate first steps towards a global understanding of the pre-NQ case, we apply the local constructions to 3-dimensional nilmanifolds carrying an abelian gerbe. These are among prime examples where T-duality is well-understood and allow us to investigate classic results in the graded language.   
\end{abstract}
\shortabstract
\begin{document}
\maketitle

\section{Introduction}
Courant algebroids play an important role in the investigation of T-duality. T-dual torus bundles with torus-invariant $H$-fluxes carry isomorphic Courant algebroids \cite{Cavalcanti:2011wu}. Originating from Buscher's observation \cite{Buscher:1987sk} this was motivated and extensively used in supergravity \cite{Grana:2008yw,Coimbra:2011nw,Coimbra:2012yy,Coimbra:2012af}. Studying dualities from the sigma-model point of view uses total spaces of Courant algebroids as target spaces \cite{Cattaneo:2009zx,Roytenberg:2006qz,Mylonas:2012pg,Chatzistavrakidis:2016jci,Chatzistavrakidis:2016jfz,Kokenyesi:2018ynq,Chatzistavrakidis:2018ztm}. The mathematical study of T-duality for principal torus bundles with $H$-flux in the geometric case and beyond showed the need for continuous fields of noncommutative and nonassociative tori \cite{Bouwknegt:2003vb,Bouwknegt:2003wp,Bouwknegt:2004ap,Bouwknegt:2003zg,Mathai:2013vua}. This is referred to as topological T-duality. In physics jargon, the latter are the nongeometric T-duals of manifolds with abelian gerbe structure. The duals of the flux re\-pre\-sen\-ting the Dixmier Douady class of the gerbe are the geometric $f$-flux and the nongeometric $Q$- and $R$-fluxes. An attempt to understand such non-geometric configurations in a still differential geometric way are the doubled spaces \cite{Hull:2006va,Hull:2007jy,Hull:2009sg} whose appearance resembles the structure of the correspondence spaces of topological T-duality. Field theory on doubled spaces, motivated by string field theory are the starting point of Double Field Theory (DFT) \cite{Tseytlin:1990va,Siegel:1993th,Siegel:1993xq,Hull:2009mi,Hull:2009zb,Hohm:2010jy,Hohm:2010pp}, which uses the momentum- and winding degrees of freedom of a closed string to parameterize its configuration space. T-duals arise as projections, sometimes called polarizations, to a physical submanifold of this doubled configuration space. The intriguing property of doubled spaces is that they allow to capture nongeometric backgrounds in a geometric way by mixing winding and momentum \cite{Hull:2007jy,Andriot:2012an,Hohm:2013bwa}. Global properties of the doubled configuration space are treated in \cite{Vaisman:2012ke,Vaisman:2012px,Papadopoulos:2014mxa,Blumenhagen:2014gva,Blumenhagen:2015zma,Marotta:2018swj,Hassler:2016srl}. The link to topological T-duality and the use of correspondence spaces however remains unclear. 

Roytenberg showed the equivalence of Courant algebroids to degree-2 symplectic differential graded ma\-ni\-folds \cite{Roytenberg:2002nu}. This gives a new view on Courant algebroid structures which turns out to be very fruitful also in case of a local description of the gauge algebra of DFT \cite{Deser:2014mxa,Deser:2016qkw,Deser:2017fko}. In particular, the C-bracket is a derived bracket on a degree-2 pre-NQ manifold and reduces to a Courant bracket after restricting to appropriate subalgebras (as is known from DFT). The aim of this proceedings contribution is to briefly review parts of the use of symplectic (pre-) NQ-manifolds in DFT and how to approach T-duality of circle bundles over tori in the presence of abelian gerbes \cite{Deser:2018flj}. The latter is a very simple case of topological T-duality and for us has the advantage that everything is well-understood and explicit enough to see how the formalism of graded manifolds applies there.

The purpose of this approach is to use different techniques in the study of duality of circle and torus bundles in order to get insight into the link between topological T-duality and DFT in the more physics oriented literature. Taking the easy case of circle bundles allows to see how to identify objects in DFT with objects used in more global descriptions of duality. 

\section{NQ-manifolds and Courant algebroids}
To begin with, our aim is to recall notions of graded linear algebra and symplectic differential graded manifolds to an extent we will need for later parts. For a detailed introduction to the subject we refer to the lecture notes \cite{Cattaneo:2010re} and the references mentioned there. To set the notation, a \emph{graded vector space} is a collection $V = (V_i)_{i\in \mathbbm{Z}}$ of finite dimensional vector spaces $V_i$\footnote{To avoid problems with tensor products we assume that the sum over all the dimensions of $V_i$ is finite.}. An element $v \in V_k$ is defined to be of degree $k$, we write $|v| = k$. For $V$ a graded vector space, the \emph{graded symmetric algebra} $S(V)$ is defined to be the quotient of the tensor algebra of $V$ by the ideal generated by elements of the form $v\otimes w - (-1)^{|v||w|} w\otimes v$. Morphisms of graded vector spaces are collections of linear maps between the respective summands. For a smooth manifold $M$, a \emph{graded manifold with body} $M$ is a locally ringed space $(M,\mathcal{O}_M)$, whose local model is
\begin{equation}\label{locmod}
  (U,\CC^\infty(U) \otimes S(V^*))\;,\qquad U\subset M\;,
\end{equation}
where $V$ is a graded vector space and $U$ is an open subset of $M$ identified with an open subset of $\mathbbm{R}^n$ if $M$ has dimension $n$. A \emph{graded vector field of degree $k$} on a graded manifold $\CM$ is a graded derivation of degree $k$, i.e.
\begin{align}
  X: \;&\CC^\infty_\bullet (\CM) \rightarrow \CC^\infty_{\bullet +k}(\CM)\;,\nonumber \\
  &X(fg) =\,X(f)g +(-1)^{k|f|}fX(g)\;.
\end{align}
We will call a degree-1 derivation $Q$ on $\CC^\infty(\CM)$ a \emph{homological vector field} if its graded commutator vanishes. A graded manifold with a homological vector field is also referred to as differential graded (dg) manifold or NQ-manifold if all generators of $\CC^\infty(\CM)$ have non-negative degree. Of particular interest for us are symplectic dg manifolds, i.e. dg manifolds carrying a 2-form\footnote{For a treatment of differential forms on graded manifolds we again refer to \cite{Cattaneo:2010re}.}, homogeneous of a certain degree (we separate form degree and the degree of the coordinates), closed with respect to the de Rham differential on the graded manifold and non-degenerate.  Many notions of ordinary symplectic geometry can be carried over to the graded setting by taking care of the grading and the local model \eqref{locmod}. In particular we have graded hamiltonian vector fields and Poisson brackets. An important class of symplectic dg manifolds are the \emph{degree-k Vinogradov algebroids} on a manifold M:
\begin{equation}
  \CV_k(M) :=T^*[k]T[1]M\;,
\end{equation}
where by $[k]$ we associate a degree $k$ to the respective fibres. This cotangent bundle is equipped with a canonical symplectic form of degree $k$. On a local patch, we have coordinates $(x^i,\xi^i,\zeta_i,p_i)$ of degrees $(0,1,k-1,k)$ such that the canonical degree-k symplectic form reads
\begin{equation}\label{Vinosymplectic}
  \omega_{\CV_k} = \, \dd x^i\wedge \dd p_i + \dd \xi^i\wedge \dd\zeta_i\;.
\end{equation}
Locally, we have the homological vector field $\lbrace \CQ_{\CV_k}, \cdot\rbrace$, where the homological function is $\CQ_{\CV_k} = \xi^ip_i$. It trivially has $\lbrace \CQ_{\CV_k}, \CQ_{\CV_k}\rbrace = 0$, but one can see in general that if a homological vector field is hamiltonian, then the Poisson square of its homological function vanishes.

We remark that degree-(k-1) functions on $\CV_k(M)$ correspond to sections of generalized versions of the tangent bundle of a manifold $M$. One can see this by looking at the local model or by writing down the most general element of degree $(k-1)$, which is
\begin{align}
  X = X^i(x)\,\zeta_i + &X_{i_1 \cdots i_{k-1}}(x)\,\xi^{i_1}\cdots\xi^{i_{k-1}}\;, \nonumber \\
  &\sim \; \Gamma(TM \oplus \wedge^{k-1}T^*M)\;.
\end{align}
This hints already towards applications in supergravity. Sections in the latter bundles are the objects of generalized geometry in case $k=2$ and for $SL(5)$-exceptional field theory in case $k=3$. The case $k=2$ turns out to be equivalent to Courant algebroids, as was shown by Roytenberg \cite{Roytenberg:2002nu} an will be the main interest in the following sections:
\begin{framed}
  {\bf $\CV_2(M)$ and Courant algebroids on $M$}\newline

  \noindent Isomorphism classes of degree-2 symplectic dg manifolds are in one-to-one correspondence with isomorphism classes of Courant algebroids.
\end{framed}
In particular, let $(E,[\cdot,\cdot]_E,\rho, \langle\cdot,\cdot\rangle)$ be a Courant algebroid over a manifold $M$ with Dorfman bracket $[\cdot,\cdot]_E$, bilinear $\langle\cdot,\cdot\rangle$ and anchor $\rho : E \rightarrow TM$. Then if $U\subset M$ is an open subset, $E|_U \simeq TU \oplus T^*U$ and sections in $E$ can be identified with degree-1 functions $X = X^i(x)\,\zeta_i + X_i(x)\, \xi^i$ on $\CV_2(U)$. Given a degree-2 NQ-manifold, using the homological function $\CQ_{\CV_2}$, one constructs the data for a Courant algebroid as follows. For smooth functions $f$ on $M$, the action of a section $\rho(X)(f)$ is obtained by $\lbrace \lbrace \CQ_{\CV_2}, X\rbrace, f\rbrace$. The Dorfman bracket of two sections $X,Y$ is $\lbrace \lbrace \CQ_{\CV_2}, X\rbrace, Y\rbrace$, i.e. they are \emph{derived brackets} using the Poisson bracket on $\CC^\infty(\CM)$ related to \eqref{Vinosymplectic}. Finally, the bilinear $\langle\cdot,\cdot\rangle$ is the Poisson bracket itself, restricted to degree-1 functions. One can now show all the axioms of a Courant algebroid. For the converse we refer to the original \cite{Roytenberg:2002nu}.

\section{Pre-NQ manifolds and Double Field Theory}\label{prenq}

\subsection{A glance at DFT}
To illustrate some key properties of closed strings which generated the construction of Double Field Theory (DFT), we consider a simple toy example of a closed string sigma model. Let $(M,G)$ be a compact Riemannian manifold of dimension $d$ with a two-form $B\in \Omega^2(M)$.  We take maps $\phi : \Sigma \rightarrow M$ for the two dimensional domain $\Sigma = \mathbbm{R} \times S^1$. Then the bosonic closed string sigma model action without dilaton is given by
\begin{equation}\label{stringsigma}
  S = \int_\Sigma \, \dd\phi^i\wedge \star_\Sigma \,\dd\phi^j\, \phi^* G_{ij} + \phi^*B\;,
\end{equation}
where $\dd$ is the differential on $\Sigma$ and $\phi^*$ denotes the pullback by $\phi$. Taking for the following $G$ and $B$ to be constant on a $d$-dimensional torus $M=T^d$ we can determine the canonical momenta and hamiltonian density for this model\footnote{For the general case we refer to the original \cite{Hull:2009mi} and the review \cite{Aldazabal:2013sca}.}. Denoting by $(\tau,\sigma)$ the coordinates on $\mathbbm{R}\times S^1$ and assume the torus periodicity $\phi^i \sim \phi^i + 2\pi$, we have
\begin{equation}
  P_i = \frac{1}{2\pi}\Bigl(G_{ij}\partial_\tau \phi^j + B_{ij}\partial_\sigma \phi^j\Bigr)\;.
\end{equation}
The hamiltonian density corresponding to \eqref{stringsigma} can be written in a suggestive form:
\begin{equation}\label{hamdensity}
  \mathfrak{h} = \frac{1}{4\pi}\Bigl(\partial_\sigma \phi, 2\pi P\Bigr)\CH(G,B)\begin{pmatrix}
    \partial_\sigma \phi\\
    2\pi P
  \end{pmatrix}\;,
\end{equation}
where $\CH$ is the \emph{generalized metric}
\begin{equation}\label{genmetric}
  \CH = \begin{pmatrix}
    \mathbbm{1} & B\\
    0 & \mathbbm{1}
  \end{pmatrix} \begin{pmatrix}
    G & 0 \\
    0 & G^{-1}
  \end{pmatrix} \begin{pmatrix}
    \mathbbm{1} & B \\
    0 & \mathbbm{1}
  \end{pmatrix}^t\;.
\end{equation}
Furthermore, periodic right- and left-moving solutions to the Euler-Lagrange equations to the sigma model \eqref{stringsigma} are given by the expansion
\begin{align}\label{stringsolution}
  \phi^k_R =& \phi^k_{0\,R} + \alpha^k_0(\tau-\sigma)  + i\underset{n\neq 0}{\sum}\,\frac{1}{n}\alpha_n^k\,e^{in(\tau-\sigma)}\;,\nonumber \\
  \phi^k_L =& \phi^k_{0\,L} + \bar \alpha^k_0(\tau + \sigma) + i\underset{n\neq 0}{\sum}\,\frac{1}{n}\bar\alpha_n^k\,e^{in(\tau + \sigma)}\;.
\end{align}
The zero modes $\phi^k_{0\,L/R}$ expressed in terms of momentum of the center of mass of the string $p_i$ and an additional quantity $w^i = \int_0^{2\pi}\partial_\sigma \phi^i \dd\sigma$ are:
\begin{equation}
\begin{aligned}
  \alpha_0^i =\,&\frac{1}{\sqrt{2}}\,G^{ij}\Bigl(p_j - (G_{jk}+B_{jk})w^k\Bigr)\;, \\
  \bar \alpha_0^i =\,&\frac{1}{\sqrt{2}}\,G^{ij}\Bigl(p_j + (G_{jk}-B_{jk})w^k\Bigr)\;.
\end{aligned}
\end{equation}
The appearance of $w^k$ on a similar footing as the center of mass momentum is the motivation to treat them as additional momenta. They are referred to as winding momenta as they are related to the winding number, e.g. of a curve parameterized by $e^{2\pi i X(\sigma)}$ in the plane example. The appearance of $w^k$ is a consequence of the solution \eqref{stringsolution} of the equations of motion to the sigma model \eqref{stringsigma} with closed string boundary conditions. The idea to associate to winding momenta an additional set of coordinates $\tilde x$ in analogy to the configuration space coordinates $x^i$ associated to the center of mass momentum $p_i$. This leads to an extension of configuration space by the winding part. In the case of closed strings on toroidal backgrounds where there are as many winding coordinates as position coordinates there is a doubling, hence the name DFT. Double fields are real or complex functions and sections in appropriate bundles depending on both sets of coordinates, i.e. locally on a direct product space $M\times \tilde M$.

Closed string theory finally provides a condition how to reduce such doubled objects to standard objects (i.e. functions, sections whose domain of definition lies in $M$). Its source is the level matching condition of closed string theory and translates into the \emph{section condition} of DFT after imposing that a product of fields satisfies the condition as soon its factors satisfy it:
\begin{equation}
  \partial_i \psi_1(x,\tilde x) \,\tilde \partial^i \psi_2(x,\tilde x) + \tilde \partial^i\psi_1(x,\tilde x)\,\partial_i \psi_2(x,\tilde x) =\,0\;.
\end{equation}
Here $\psi_i$ denote ``any kind of field'' appearing in the theory and $\tilde \partial^i$ is the partial derivative with respect to the winding coordinates. One aim of the later sections is to understand this condition better as it is e.g. not clear how it is meant on vector fields or tensors like the metric.

A great success of DFT is to provide an action principle for the field $\CH(G,B)$, appearing already in \eqref{genmetric}, but now promoted to depend on the doubled coordinate set on $M\times \tilde M$. The action reduces correctly to the NSNS sector of standard type IIA supergravity by dropping the winding coordinates (the easiest solution of the section condition) and thus provides an interesting generalization. It is given by
\begin{align}\label{DFTaction}
  S(\CH) =\, &\int_{M\times \tilde M}\,\dd x\,\dd \tilde x\, e^{-2\CD}\Bigl(\frac{1}{8}\,\CH^{MN}\partial_M\CH^{KL}\partial_N\CH_{KL} \nonumber \\
  &- \frac{1}{2}\,\CH^{MN}\partial_N\CH^{KL}\partial_L \CH_{MK} +2\,\partial_M\CD \,\partial_N\CH^{MN} \nonumber \\
  &+ 4\,\CH^{MN}\partial_M\CD\partial_N \CD\Bigr)\;.
\end{align}
The notation here is as follows: Objects with capital indices are defined to transform in the fundamental representation of $O(d,d)$ ($d$ is the dimension of $M$), the matrix group leaving the corresponding bilinear form $\eta$ invariant:
\begin{equation}
  A\in \textrm{Mat}(2d,\mathbbm{R})\;,\quad A^t \eta A = \,\eta\;,\qquad \eta = \begin{pmatrix}
    0 & \mathbbm{1}\\
    \mathbbm{1} & 0
  \end{pmatrix}
\end{equation}
Double functions are defined to be $O(d,d)$ invariant functions on $M\times \tilde M$, an example being $\CD$ in \eqref{DFTaction}. Double vectors are objects $\CX^M(x,\tilde x) = (\CX^i(x,\tilde x),\CX_i(x,\tilde x))$ and the double metric $\CH$ is a symmetric two-tensor under $O(d,d)$ depending on $G(x,\tilde x)$ and $B(x,\tilde x)$ in the form \eqref{genmetric}. Capital indices are raised and lowered with $\eta$, e.g. $\partial^M = \eta^{MN}\partial_N$.

In addition to its manifest $O(d,d,\mathbbm{R})$-invariance, the action \eqref{DFTaction} has a local gauge symmetry parameterized by doubled vectors. The gauge transformations have a suggestive form which is the motivation to locally introduce a double version of the Lie bracket $\CL_\CX$. Its action on double functions $\psi$  and vectors $\CY$ is
\begin{align}
  \CL_\CX \,\psi =\,&\CX^M\partial_M \psi\;,\nonumber \\
  (\CL_\CX \,\CY)^M =\,&\CX^K\partial_K \CY^M  - \Bigl(\partial_K \CX^M - \partial^M \CX_K\Bigr)\CY^K\;,
\end{align}
and extended to tensor products by the Leibniz rule. The commutator of such gauge transformations closes and defines the \emph{C-bracket} of two double vectors:
\begin{equation}
  [\CL_\CX , \CL_\CY] =\; \CL_{[\CX,\CY]_C}\;,
\end{equation}
where the component form of the C-bracket is 
\begin{align}\label{Cbracket}
  [\CX,\CY]_C^M =\,\CX^K&\partial_K \CY^M - \CY^K\partial_K \CX^M \nonumber \\
  &+ \frac{1}{2}\Bigl(\CY_K\partial^M \CX^K - \CX_K\partial^M \CY^K\Bigr)\;.
\end{align}
The C-bracket has properties remarkably similar to the Courant bracket in generalized geometry and indeed reduces to it by solving the strong constraint (e.g. in the form of $\tilde \partial^i = 0$). However it cannot be simply the Courant bracket on the cartesian product $M\times \tilde M$ as already sections in the generalized tangent bundle on $M\times \tilde M$ would have too many components. It as well cannot be a bracket on some Dirac structure of a Courant algebroid on $M\times \tilde M$ as this would be a Lie algebroid and therefore its bracket would satisfy the Jacobi identity which is not the case for \eqref{Cbracket}. In the next section, our main aim is to find out what an appropriate viewpoint on the C-bracket can be and clarify its close resemblance to Courant algebroid structures.

\subsection{Local aspects of the gauge algebra of DFT in NQ-language}
The idea is to apply a derived bracket construction similar to Roytenberg's for Courant algebroids in order to understand the objects of DFT and in particular the bracket \eqref{Cbracket}. Locally, we model the doubling of $d$-dimensional configuration space $M$ by taking the direct product of two copies $K := M_x \times \tilde M_{\tilde x}$. The naive idea to write down the degree-2 symplectic dg manifold corresponding to the standard Courant algebroid over $K$ gives too many vectors, i.e. degree-1 functions: for $x^M = (x^\mu,\tilde x_\mu)$\footnote{A motivation for the index structure is the case $K=T^*M$ for phase space models, see e.g. \cite{Aschieri:2015roa,Deser:2016qkw}}, we get
\begin{equation}
  \CC^\infty(\CV_2(K)) \ni X = X^M(x,\tilde x)\zeta_M + X_M(x,\tilde x)\xi^M\;,
\end{equation}
where e.g. $\zeta_M = (\zeta_\mu, \zeta^\mu)$ and $\zeta^\mu$ correspond to the tangent vectors along the $\tilde x_\mu$ directions. Taking $\CV_2(K)$ with its canonical symplectic structure
\begin{equation}\label{sympK}
  \omega_{\CV_2(K)} = \dd x^M \wedge \dd p_M + \dd\xi^M\wedge \dd\zeta_M\;,
\end{equation}
we reduce the number of degree-one generators by first defining
\begin{equation}
  \theta^M_\pm =\,\frac{1}{2}\Bigl(\xi^M \pm \eta^{MN}\zeta_N\Bigr)\;,
\end{equation}
and taking the quotient of $\CC^\infty(\CV_2(K))$ by the ideal generated by the $\theta^M_-$. We use the $O(d,d)$-invariant $\eta$, which arises naturally in the Poisson brackets of the $\xi^M, \zeta_M$\footnote{Note, that due to the degree of $\xi^M$ and $\zeta_M$ and the degree-2 symplectic form \eqref{sympK}, the Poisson brackets are symmetric, so we get $O(d,d)$ rather than $Sp(2d)$.}. Hence we arrive at a symplectic graded manifold $\CE_K$ whose algebra of functions is generated by $(x^M, \theta_M,p_M)$ of degree $(0,1,2)$, with symplectic structure\footnote{We raise and lower capital indices with $\eta$, e.g. $\theta^M = \eta^{MN}\theta_N$.}
\begin{equation}\label{sympE}
  \omega_{\CE_K} = \dd x^M \wedge \dd p_M + \tfrac{1}{2}\eta^{MN}\, \dd\theta_M\wedge \dd\theta_N\;.
\end{equation}
The homological function on $\CV_2(K)$ reduces to a degree-3 function $\CQ_{\CE_K}$ on $\CE_K$ whose square is not zero any more, but whose hamiltonian vector field preserves the symplectic structure \eqref{sympE}
\begin{equation}\label{homfunktionE}
  \CQ_{\CE_K} = \theta^M p_M\;,\quad \lbrace \CQ_{\CE_K}, \CQ_{\CE_K}\rbrace = \eta^{MN}p_Mp_N\;.
\end{equation}
In general, we want to call a symplectic graded manifold (of degree $k$) with a function of degree $(k+1)$ whose hamiltonian vector field preserves the symplectic structure a \emph{pre-NQ manifold}. So $\CE_K$ is a degree-2 pre-NQ manifold. In degree two, such manifolds have the possibility to find subalgebras of their algebras of functions which allow two-term $L_\infty$-structures. We refer the reader to \cite{Lada:1992wc,Lada:1994mn,Barnich:1997ij,Stasheff:2018vnl} and the references listed there for an introduction and history of $L_\infty$ structures. Let $\CM$ be a pre-NQ manifold of degree 2 with pre-homological function $\CQ$. On $\CC^\infty_1(\CM) \oplus \CC^\infty_0(\CM)$ we define the maps
\begin{align}
  \mu_1 &: \quad \CC_0^\infty(\CM) \rightarrow \CC^\infty_1(\CM)\;,\nonumber  \\
    \mu_2 &: \quad \CC_i^\infty(\CM) \times \CC_j^\infty(\CM) \rightarrow \CC_{i+j-1}^\infty(\CM) \quad i+j >0\;,\nonumber \\
    \mu_3 &: \quad \CC_i^\infty(\CM) \times \CC^\infty_1(\CM) \times \CC^\infty_1(\CM) \rightarrow \CC^\infty_0(\CM) \quad i=0,1\;.
\end{align}
With $\delta f := \lbrace \CQ,f\rbrace$ for $f\in \CC_1^\infty(\CM)$ and zero otherwise, they are
\begin{align}
  \mu_1(f) &=\; \lbrace \CQ,f\rbrace \;,\nonumber \\
    \mu_2(f,g) &=\; \tfrac{1}{2}\Bigl(\lbrace \delta f, g\rbrace \pm \lbrace \delta g, f \rbrace\Bigr)\;, \nonumber \\
    \mu_3(f,g,h)&=\; -\tfrac{1}{12}\Bigl(\Bigl\lbrace \lbrace \delta f,g\rbrace,h\Bigr\rbrace \pm \dots\Bigr)\;,
\end{align}
where the dots denote a sum over graded permutations of $\lbrace f,g,h\rbrace$.
Having defined these maps, it is now possible to find conditions when a subalgebra of $\CC^\infty(\CM)$, concentrated in degrees $0,1$ forms a 2-term $L_\infty$ algebra. We use $Q^2$ for $\lbrace \CQ,\lbrace \CQ,\cdot\rbrace\rbrace$.
\begin{framed}
  Consider a subset of $\CC^\infty_1(\CM) \oplus \CC^\infty_0(\CM)$, such that the Poisson brackets and the maps $\mu_i$ close on this subset. Then the latter is a 2-term $L_\infty$-algebra if and only if
    \begin{align}\label{Lie2cond}
      \lbrace Q^2 f, g\rbrace + \lbrace Q^2 g, f\rbrace =&0\,,\nonumber \\
      \lbrace Q^2 X,f\rbrace + \lbrace Q^2 f, X\rbrace =&0\,,\nonumber \\
      \lbrace \lbrace Q^2X,Y\rbrace,Z\rbrace_{[X,Y,Z]} =&0\,, 
    \end{align}
    for functions $f,g$ of degree 0 and $X,Y,Z$ of degree 1. The subscript $[X,Y,Z]$ means the alternating sum over $X,Y,Z$.
\end{framed}
We will now apply this to our local description of DFT. Using the Poisson brackets associated to \eqref{sympE} and the pre-homological function \eqref{homfunktionE} we first notice the result for the maps $\mu_i$. For a degree-0 function $f$, degree-1 functions $\CX = \CX^M(x,\tilde x) \theta_M$, $\CY = \CY^M(x,\tilde x) \theta_M$ we get
\begin{align}
  \mu_1(f) =\,&\theta^M\partial_M f \;,\nonumber \\
  \mu_2(\CX,\CY) =\, &\tfrac{1}{2}[\CX,\CY]_C^M \theta_M\;, \nonumber \\
  \mu_3(\CX,\CY,\CZ) =\,&\CX^M\CZ^N\partial_M\CY_N - \CY^M\CZ^N\partial_M\CX_N \nonumber \\
  &+ \textrm{cycl.}_{\CX\CY\CZ}\;.
\end{align}
Thus, $\mu_1$ gives the analogue of the de Rham differential on the doubled space, $\mu_2$ the C-bracket \eqref{Cbracket} and the expression for $\mu_3$ gives the failure of the C-bracket's Jacobiator to vanish. Furthermore, we can evaluate the conditions \eqref{Lie2cond} whose solution gives subalgebras with 2-term $L_\infty$ structures, i.e. Courant algebroids. Denoting by $Q_{\CE_K}^2$ the action of $\lbrace \CQ_{\CE_K},\lbrace \CQ_{\CE_K},\cdot\rbrace \rbrace $, on degree-0 functions $f$ and degree-1 functions $\CX,\CY,\CZ$ as above we get:
\begin{align}
  &\lbrace Q_{\CE_K}^2 f, g\rbrace + \lbrace Q_{\CE_K}^2 g, f\rbrace = 2\partial_M f \,\partial^M g\;,\nonumber \\
  &\lbrace Q_{\CE_K}^2 \CX, f \rbrace + \lbrace Q_{\CE_K}^2 f,\CX\rbrace = 2\partial_M \CX\,\partial^M f \;,\nonumber \\
  &\lbrace \lbrace Q^2\CX,\CY\rbrace ,\CZ\rbrace_{[\CX,\CY,\CZ]} = \nonumber\\
  &\hspace{40pt} 2\theta^L\Bigl((\partial^M \CX_L)(\partial_M \CY^K)\CZ_K\Bigr)_{[\CX,\CY,\CZ]}\;.
\end{align}
We see that the first two expressions correspond to the section condition of DFT on two functions and on a function and a vector. The third expression turns out to be important in the investigation of Riemann curvature operators in the setting of degree-2 pre-NQ manifolds, we refer to \cite{Deser:2016qkw} for details.

To sum up, we locally constructed a setting which allows to identify the C-bracket of DFT as a derived bracket in a similar way as Roytenberg identified the Courant bracket as a derived bracket. DFT's section condition can be algebraically interpreted as the obstructions for a subalgebra of functions concentrated in degree 1 and 2 of a pre-NQ manifold to form a 2-term $L_\infty$-algebra. This allows for some algebraic insight into the structures relevant for DFT. Of course there is much more to do. The action principle of DFT uses integration on doubled configuration spaces and a Ricci-scalar for a generalized tangent bundle which we identified as a pre-NQ-manifold. Understanding Riemannian geometry for pre-NQ-manifolds and therefore DFT dynamics is a next goal and first steps have been performed in \cite{Deser:2016qkw}. Rather than going into this, in the next sections we will show that apart from the local considerations so far, global and therefore more geo\-me\-tric statements can be obtained, at least for simple examples relevant for physics (and in principle in general). Instead of considering a simple direct product $M\times \tilde M$ we will use fibred products over some base manifolds, which are referred to as correspondence spaces. It turns out that our construction of DFT objects is also possible there giving some first insights into global structures in DFT and into the geometry of pre-NQ manifolds.

\section{3-dimensional nilmanifolds and correspondence spaces}
We now specialize to an explicit class of examples, the case of 3-dimensional Heisenberg nilmanifolds \cite{eberlein1994geometry},\linebreak equipped with an $H$-flux or abelian gerbe. These are circle bundles over the 2-torus and the $H$-flux twists the Courant algebroid on their generalized tangent space. The geometry, as well as T-duality of circle bundles, is very well-understood \cite{Bouwknegt:2003vb,Bouwknegt:2003wp} which has the advantage that we have a clear and explicit starting point. The idea is to reformulate the well-known results for this case in our language in order to identify the objects relevant for duality. This will be the starting point to explore more complicated examples of doubled spaces in our language in future work.

\subsection{Lightning review of the setting}
Consider a 3-dimensional nilmanifold in the form of an $S^1$-bundle over the 2-torus. The bundle structure can be described either with local transition functions or periodicity conditions as this type of geometry arises as a quotient of $\mathbbm{R}^3$. We prefer the latter as it is convenient for moving from the local description of the previous section to a more global viewpoint. The equivalence relation imposed on flat space is
\begin{align}\label{period}
  (x^1,x^2,x^3) &\sim (x^1,x^2+1,x^3) \sim (x^1,x^2,x^3+1)\nonumber \\
  &\sim (x^1+1,x^2,x^3-jx^2)\;.
\end{align}
This gives an $S^1$-bundle $\pi_j : N_j \rightarrow T^2$ with base coordinates $x^1,x^2$. A basis of globally defined one-forms is $\lbrace \dd x^1, \dd x^2, \dd x^3 + jx^1\,\dd x^2\rbrace $ and the third defines a connection $A = \,\dd x^3 + jx^1 \dd x^2$ on the bundle with first Chern class $c_1(N_j) = \dd A = j\dd x^1\wedge \dd x^2$. Finally we introduce an abelian gerbe on $N_j$ by giving a two-form $B \in \Omega^2(N_j)$ satisfying
\begin{equation}\label{gerbe}
  B\vert _{U_b} =\, B\vert_{U_a} + \,\dd\alpha\vert_{U_a \cap U_b}\; \quad \textrm{on}\;\;\; U_a \cap U_b\;,
\end{equation}
for open subsets $U_a, U_b \subset N_j$ and $\alpha \in \Omega^1(U_a\cap U_b)$. Choose $B = k\,x^1\,\dd x^2\wedge \dd x^3$, then the only non-trivial relation \eqref{gerbe} is for the $x^1$-periodicity and reads
\begin{equation}
  B\vert_{x^1+1} - B\vert_{x^1} =\, k\, \dd x^2\wedge \dd x^3 = \dd\alpha\;, \quad \alpha = \, kx^2\dd x^3\;.
\end{equation}
In total, the geometric setup is determined by two integers, the integral over the first Chern class of the $S^1$-bundle and the integral over the globally defined $H$-flux $H=\dd B$, representing the Dixmier Douady class of the gerbe:
\begin{equation}
  j =\; \int_{T^2} \, c_1(N_j) \;, \quad k =\; \int_{N_j} \, H \;.
\end{equation}
We call this setup therefore $N_{j,k}$.

\subsection{Review of T-duality for circle bundles}
Following \cite{Bouwknegt:2003vb,Bouwknegt:2003wp}, T-duality for circle bundles is defined in general by the following construction. Let $P \stackrel{\pi}{\rightarrow} M$ and $\tilde P \stackrel{\tilde \pi}{\rightarrow} M $ be two $S^1$-bundles over the common connected and compact base $M$. Assume that $H$ and $\tilde H$ are closed, integral 3-forms on $P$ and $\tilde P$, such that they are mapped to closed, integral two-forms on the base after integration along the respective $S^1$-fibres. Let $\CCG$ and $\tilde \CCG $ denote the corresponding abelian gerbes whose Dixmier Douady classes are represented by $H$ and $\tilde H$. To define T-duality, consider the correspondence space which is $P\times_M \tilde P$, the fibred product of $P$ and $\tilde P$ over the base, i.e. the following commutative diagram:
\begin{equation}\label{eq:generic_diagram}
 \xymatrix{
     & & P\times_{M}\tilde P \ar@{->}[dl]_{\pr} \ar@{->}[dr]^{\tilde \pr}& & \\
     \CCG \ar@{->}[r] & P\ar@{->}[dr]_{\pi} & & \tilde P\ar@{->}[dl]^{\tilde \pi} & \tilde \CCG\ar@{->}[l] \\
     & & M & &
    }
\end{equation}
In this situation, the circle bundle $\tilde P$ can be chosen such that its first Chern class is given by the flux $H$ integrated along the $S^1$ fibre. Due to the classification of circle bundles by their first Chern class, this is possible in a unique way. Using a Gysin sequence argument, \cite{Bouwknegt:2003vb} show that an $\tilde H$-flux can be chosen in such a way that
\begin{equation}\label{dualitycond}
  \tilde \pr^* \tilde H - \pr^* H = \, \dd (\pr^* A \wedge \tilde \pr^* \tilde A) =:\dd\CF \;,
\end{equation}
where $\tilde A$ denotes the connection on the bundle $\tilde P$. Condition \eqref{dualitycond} is sometimes used as a \emph{definition} for two circle (or more general torus-) bundles $P$ and $\tilde P$ to be T-dual \cite{Cavalcanti:2011wu}. It means that T-dual circle bundles have cohomologous $H$-fluxes on the correspondence space. An example are pairs of nilmanifolds $N_{j,k}$ and $N_{k,j}$ with integers of first Chern class and Dixmier Douady class interchanged. Denoting by $\tilde \pr_* : P\times_M \tilde P \rightarrow \tilde P$ the fibre integration along the $S^1$ fibre of $P$, \cite{Bouwknegt:2003vb} define the \emph{T-duality map} to be the homomorphism $\tilde \pr_* \circ e^{\CF \wedge}\circ \pr^*$. It allows to map closed forms on $P$ with respect to the $H$-twisted differential $\dd+ H\wedge$ to closed forms on $\tilde P$ with respect to the $\tilde H$-twisted differential and descends to an isomorphism on the corresponding twisted cohomologies. The aim of the next section is to understand these results in terms of the NQ-language.

\section{T-duality of nilmanifolds in the pre-NQ language}
The local results of section \ref{prenq} can be patched together by correctly taking into account the non-trivial gerbe structures of the nilmanifolds $N_{j,k}$ and $N_{k,j}$. It is most convenient to use the description of these objects in terms of the twisted periodicity conditions \eqref{period}. The nontrivial  gerbes affect the periodicity of the degree-1 coordinates and hence both the bundle- and gerbe- contributions affect the periodicity of the degree-2 momenta.
\subsection{Gerbes on nilmanifolds and periodicity}
Taking $N_{j,k}$ with periodicity \eqref{period}, we introduce coordinates on $\CCC_{j,k} = T^*[2]T[1]N_{j,k}$ as follows. First, consider the case $k=0$. Degree-1 coordinates correspond to the global basis of one-forms and their linear duals:
\begin{equation}\label{globalforms}
\begin{aligned}
  \bar \xi^1 &\sim \dd x^1\;, \quad \bar \xi^2 \sim \dd x^2\;,\quad \bar \xi^3 \sim \dd x^3 + jx^1 \,\dd x^2\;,\\
  \bar \zeta_1 &\sim \partial_{x^1} \quad \bar \zeta_2 \sim \partial_{x^2}-jx^1\,\partial_{x^3}\, \quad \bar \zeta_3 \sim \partial_{x^3}\;.
\end{aligned}
\end{equation}
The bar is used to distinguish from another set of local coordinates to be introduced soon which also takes into account the gerbe structure. The coordinates \eqref{globalforms} have the advantage that they are global as they come from globally defined objects. The degree-2 momenta are the standard ones associated to the $x^i$ by taking the cotangent $T^*[2]$. Their periodicity is inferred from the periodicity properties of local coordinates on $T[1]N_{j,k}$ and were given e.g. in \cite{Roytenberg:2002nu}: For a coordinate change $x^{i'} = x^{i'}(x^i)$ on a manifold $M$ and corresponding change $e_{a'} = \,T^a_{a'}(x^i) e_a$ of a local basis of sections in $TM$, the momenta transform as
\begin{equation}\label{ptransform}
  p_{i} =\, \frac{\partial x^{i'}(x)}{\partial x^i}\,p_{i'} + \frac{1}{2}\,\xi^{a'}\frac{\partial T^a_{a'}(x)}{\partial x^i} \,g_{ab} T^b_{b'}(x)\,\xi^{b'}\;,
\end{equation}
where $g_{ab}$ is inferred from the isomorphism relating the chosen basis of one-forms to the corresponding basis on $TM$. In our example, due to the $\bar \xi^i$ and $\bar \zeta_i$ being global, the periodicity of the momenta is a consequence of \eqref{period} and reads
\begin{equation}
  (\bar p_1, \bar p_2, \bar p_3)\vert_{x^1+1} = (\bar p_1, \bar p_2 + j\,\bar p_3, \bar p_3)\;,
\end{equation}
so that in total we have the following periodicity for local coordinates on $\CCC_{j,k}$
\begin{align}
  &(x^1,x^2,x^3,\bar \xi^1, \bar \xi^2, \bar \xi^3, \bar \zeta_1, \bar \zeta_2, \bar \zeta_3, \bar p_1, \bar p_2, \bar p_3)\nonumber \\
  \sim &(x^1+1, x^2, x^3-jx^2,\bar \xi^1, \bar \xi^2, \bar \xi^3, \bar \zeta_1, \bar \zeta_2 \bar \zeta_3, \bar p_1, \bar p_2 + j\bar p_3, \bar p_3)\;.
\end{align}
To complete the data for $\CCC_{j,k}$, the symplectic form and homological function are given by
\begin{align}
  \omega =& \,\dd x^i \wedge \dd\bar p_i + \dd\bar \xi^i \wedge \dd\bar \zeta_i \nonumber\\
  \CQ_0 =& \bar \xi^i \bar p_i - jx^1 \bar \xi^2 \bar p_3\;.
\end{align}
The second part of the homological function can be understood by the contribution of the non-trivial anchor map relating the $\bar \zeta$-basis to the standard basis of sections on the tangent bundle. Phrased differently, it contains the structure constants of the non-trivial Lie bracket. Both, $\omega$ and $\CQ_0$ are invariant under periodicity and hence are well-defined and global on $\CCC_{j,k}$.

Next, we include the non-trivial gerbe structure, i.e. the case $k\neq 0$ in our example. For a manifold $M$ with open cover $\sqcup_a U_a \twoheadrightarrow M$, a \emph{connective structure} is given by a collection of one-forms $\alpha_{(ab)}$ on intersections $U_a \cap U_b$ such that $\alpha_{(ab)} = -\alpha_{(ba)}$ and with the condition
\begin{equation}
  \alpha_{(ab)} + \alpha_{(bc)} + \alpha_{(ca)} = g^{-1}_{(abc)}\,\dd g_{(abc)}\;,
\end{equation}
where $g_{(abc)}$ form a 2-cocycle $g$ representing an element in $H^2(M,U(1))$ with
\begin{equation}
  (\delta g)_{(abcd)} = g_{(bcd)}g^{-1}_{(acd)} g_{(abd)} g^{-1}_{(abc)} = 1 \;,
\end{equation}
where $\delta$ is the corresponding coboundary map\footnote{Thus, $\dd\alpha_{(ab)} + \dd\alpha_{(bc)} + \dd\alpha_{(ca)} = 0\;, \quad \textrm{on} \;\;\;U_a \cap U_b \cap U_c\;$}. A \emph{generalized tangent bundle twisted by a gerbe} is then determined by a collection of sections of the generalized tangent bundle $\lbrace X_{(a)} + \gamma_{(a)} \in \Gamma(TU_a \oplus T^*U_a)\rbrace$ such that on intersections $U_a \cap U_b$
\begin{equation}\label{gerbetrafo}
  X_{(a)} + \gamma_{(a)} = \, X_{(b)} + \gamma_{(b)} + \iota_{X_{(b)}} \dd\alpha_{(ab)}\;.
  \end{equation}
Denoting the curving $B \in \Omega^2(M)\otimes U(1)$, such that $B_{(a)} - B_{(b)} = \dd\alpha_{(ab)}$, it is easy to see that one can define sections in the generalized tangent bundle invariant under \eqref{gerbetrafo} by $e^B(X+\gamma) := X+\gamma +\iota_X B$.

Specifying to our example, in the case of nonvanishing $k$, the coordinates $\bar \xi^i, \bar \zeta_i$ are not invariant any more due to \eqref{gerbetrafo}. Hence we take their invariant versions, $(\xi, \zeta) := e^{\lbrace \cdot, B\rbrace}(\bar \xi, \bar \zeta)$. This means that we change to the following set of coordinates, taking the $x^i$ as usual
\begin{equation}\label{gerbecoord}
  \xi^i = \bar \xi^i\;, \quad \zeta_1 = \bar \zeta_1 \;,\quad \zeta_2 = \bar \zeta_2 - kx^1\,\bar \xi^3\; \quad
  \zeta_3 = \bar \zeta_3 + kx^1\,\bar \xi^2\;.
\end{equation}
Due to \eqref{ptransform}, this change also leads to a change in the momenta:
\begin{equation}
  p_1 = \bar p_1 - k\bar \xi^2\bar \xi^3\;,\quad p_2 = \bar p_2\;,\quad p_3 = \bar p_3\;.
\end{equation}
As one can check, this coordinate change is a symplectomorphism. The symplectic form $\omega$ does not change. The homological function receives a contribution due to the $H$-flux as expected. Furthermore the periodicity of the variables stays the same as the base coordinates did not change and the new $(\xi,\zeta)$ are invariant. We summarize our description in the following.
\begin{framed}
  \paragraph{The Courant algebroid $\CCC_{j,k}$} 
\begin{center}
  Periodicity:
  \begin{align}\label{finalcoord}
    &(x^1,x^2,x^3,\xi^i,\zeta_i,p_1,p_2,p_3) \nonumber \\
    \sim &(x^1+1,x^2,x^3 -jx^2,\xi^i,\zeta_i,p_1,p_2 + jp_3, p_3)\;.
  \end{align}
  Symplectic form:
  \begin{equation}
    \omega = \,\dd x^i \wedge \dd p_i + \dd \xi^i \wedge \dd\zeta_i\;.
  \end{equation}
  Homological function:
  \begin{equation}
    \CQ = \, \xi^ip_i -jx^1 \xi^2 p_3 + k\xi^1\xi^1\xi^3\;.
  \end{equation}
\end{center}
\end{framed}

We remark that $\omega$ and $\CQ$ are invariant under periodicity and hence well-defined and global. The $k$-dependent term in the homological function gives the $H$-twisted Courant bracket after the derived bracket construction.

\subsection{The correspondence space as pre-NQ manifold}
We describe the correspondence space $K:= N_{j,k}\times_{T^2} N_{k,j}$ in terms of periodicity conditions. The base components are the same as for the base in \eqref{finalcoord}. As $K$ is at the same time a circle bundle over $N_{j,k}$ with first Chern number $k$ and a circle bundle over $N_{k,j}$ with first Chern number $j$, we use the following periodicity:
\begin{align}
  &(x^1,x^2,x^3,x^4)\sim (x^1,x^2+1,x^3,x^4)\sim (x^1,x^2,x^3+1,x^4) \nonumber \\
  \sim &(x^1,x^2,x^3,x^4+1)\sim (x^1+1,x^2,x^3 -jx^2,x^4-kx^2)\;.
\end{align}
Next we follow the program outlined in section \ref{prenq}. First we write down the Courant algebroid structure on $K$ using $T^*[2]T[1]K$, twisted by the 3-form $\pr^* H + \tilde \pr^* \tilde H$. We then reduce the number of degree-1 variables of the $S^1$-fibre coordinates\footnote{These are the ones to be compared with section \ref{prenq}, as they appear as ``doubled'' in the correspondence space. Phrased differently, the $S^1$ fibre of $N_{j,k}$ gets doubled by adding the $S^1$-fibre of $N_{k,j}$.} to the subset constructed in section \ref{prenq}. This gives the pre-NQ manifold relevant for the double field description of $N_{j,k}$.
The Courant algebroid $\CCC_K$ on $K$ is given in a similar way as in the previous section for $N_{j,k}$, so we only  give the result: The periodicity of the degree-1 and degree-2 coordinates is
\begin{align}
  &(x^1,\xi^1,\xi^2,\xi^3,\xi^4,\zeta_1,\zeta_2,\zeta_3,\zeta_4,p_1,p_2,p_3,p_4)\nonumber \\
  \sim & (x^1+1,\xi^1,\xi^2,\xi^3,\xi^4,\zeta_1,\zeta_2,\zeta_3,\zeta_4,\nonumber \\
  & \hspace{70pt} p_1,p_2+jp_3 + kp_4, p_3,p_4)\;.
\end{align}
The symplectic structure is in Darboux form, we use indices $\mu, \nu \in \lbrace 1,\dots, 4\rbrace$ on the correspondence space K:
\begin{equation}\label{sympCK}
  \omega_{\CC_K} = \; \dd x^\mu \wedge \dd p_\mu + \dd\xi^\mu \wedge \dd\zeta_\mu\;,
\end{equation}
and the homological function contains the structure constants encoding the two circle bundles as well as the twists by the gerbe structures:
\begin{align}\label{QCK}
  \CQ_{\CC_K} =\,\xi^\mu p_\mu -jx^1\xi^2&p_3 - kx^1\xi^2p_4 \nonumber \\
  &+ k\,\xi^1\xi^2\xi^3 + j\,\xi^1\xi^2\xi^4\;.
\end{align}
The homological function is symmetric under the exchange of $(j,k)$ and $(x^3,x^4)$ respectively, due to the construction of $K$ from nilmanifolds with exchanged integral numbers of first Chern class and Dixmier Douady class.

Next, we construct the pre-NQ manifold $\CCE_K$ with body $K$ allowing to state features of Double Field Theory for our example in a global way. For this we define
\begin{equation}
  \theta^3_\pm :=\,\frac{1}{2}\Bigl(\xi^3 \pm \zeta_4\Bigr)\;,\quad \theta^4_\pm := \, \Bigl(\xi^4\pm \zeta_3\Bigr)\;,
\end{equation}
and drop the coordinates $\theta^i_- ,\;i\in \lbrace 3, 4\rbrace$. Phrased differently the algebra $\CC^\infty(\CCE_K)$ is obtained by $\CC^\infty(\CCC_K)/\CI$ where $\CI$ is the ideal generated by $\theta_-^3$ and $\theta_-^4$. The resulting $\CCE_K$ is a symplectic graded manifold with degree-0 part $K$. It's symplectic structure is \eqref{sympCK}, reduced to $\CC^\infty(\CCE_K)$. The homological function \eqref{QCK} is reduced in a similar way to a function $\CQ_{\CCE_K}$. Note that the latter does not square to zero any more. Rather, its square provides us with an algebraic form of the strong section condition which we will investigate later in the section. We will therefore call this function ``pre-homological'', also because it will reduce to homological functions after projecting to the Courant algebroids on $N_{j,k}$ and $N_{k,j}$. We summarize:
\begin{framed}
  \paragraph{The pre-NQ manifold $\CCE_K$}
  \begin{center}
    Periodicity:
    \begin{align}\label{Eperiod}
      &(x^1,\xi^1,\xi^2,\zeta_1,\zeta_2,\theta_+^3,\theta_+^4,p_1,p_2,p_3,p_4) \nonumber \\
      \sim &(x^1+1,\xi^1,\xi^2,\zeta_1,\zeta_2,\theta_+^3,\theta_+^4,\nonumber \\
      &\hspace{40pt} p_1,p_2+jp_3+kp_4,p_3,p_4)\;.
    \end{align}
    Symplectic form:
    \begin{align}\label{Esymplectic}
      \omega_{\CCE_K}= \dd x^\mu\wedge \dd p_\mu &+ \dd\xi^1\wedge \dd\zeta_1 + \dd\xi^2\wedge \dd\zeta_2 \nonumber \\
      &+ \dd\theta_+^3\wedge \dd\theta_+ ^4\;.
    \end{align}
    ``Pre-homological'' function:
    \begin{align}\label{prehom}
      \CQ_{\CCE_K}=\; \xi^1p_1& + \xi^2p_2 + \theta_+^3  p_3 + \theta_+^4 p_4 \nonumber \\
      &-jx^1\xi^2p_3 - kx^1\xi^2p_4 \nonumber \\
      &+ k\,\xi^1\xi^2\theta_+^3 + j\,\xi^1\xi^2\theta_+^4\;.
    \end{align}
  \end{center}
\end{framed}

A few remarks are in order. First, $\CCE_K$ is pre-NQ as the hamiltonian vector field $Q_{\CCE_K} = \lbrace \CQ_{\CCE_K},\cdot\rbrace$ preserves the symplectic form: $\mathcal{L}_{Q_{\CCE_K}} \omega_{\CCE_K} = 0$. Second, the Poisson square of $\CQ_{\CCE_K}$ reduces to a simple form:
\begin{equation}\label{easyfactor}
  \lbrace \CQ_{\CCE_K},\CQ_{\CCE_K}\rbrace = \; 2\,p_3 p_4\;.
\end{equation}
This factorization will be important later on. It is the source for finding solutions to the section condition which allows in turn to find projections from $\CE_K$ to the Courant algebroids on $N_{j,k}$ and $N_{k,j}$. Of course these were known from before, but as emphasized in the beginning, this is to see how these projections can be found and serves as a starting point to do a similar procedure for cases where the projections are not known a priori. Finally, the factorization in the simple form \eqref{easyfactor} is due to the fact that $(j,k)$ appear interchanged in $N_{j,k}$ and $N_{k,j}$. Up to this stage the construction can be performed for any pair of nilmanifolds $N_{j,k}$ and $N_{m,n}$. After determining the correspondence space and reducing to $\CE_K$, the Poisson square of the pre-homological functions becomes
\begin{equation}\label{alg}
  \lbrace \CQ_{\CCE_K},\CQ_{\CCE_K}\rbrace = \,2(p_3 + (k-m)\xi^1\xi^2)(p_4 + (n-j)\xi^1\xi^2)\;.
\end{equation}
As the coordinates were chosen in a way that the pre-homological functions are invariant under periodicity and therefore global, the right hand side is well-defined on the correspondence space. We leave a geometric interpretation of the result for future work. 
    
\subsection{T-duality of circle bundles reformulated}
The most direct way to solve the section condition is to define projections from $\CC^\infty(\CE_K)$ to subalgebras given by the kernels of the derivations $D_3 := \lbrace p_3,\cdot\rbrace$ and $D_4 =\lbrace p_4,\cdot\rbrace$. These determine $L_\infty$-algebras, as was shown in detail in \cite{Deser:2018flj}. Geometrically they correspond to projections constant along the fibre directions $\partial_{x^3}$ or $\partial_{x^4}$. This is the analogue of setting either position- or winding-degrees of freedom to zero in Double Field Theory. Locally, we also want the projections to be constant along the corresponding conjugate momenta. We want to achieve this globally in our example and there are two natural candidates of vector fields determining kernels of projections together with $D_3$ and $D_4$. Let $\CA_3$ and $\CA_4$ be the one-forms on $\CCE_K$ which are lifts of the connection forms $\pr^* A = \dd x^3 + jx^1\,\dd x^2$ and $\tilde \pr^* \tilde A = \dd x^4+kx^1\,\dd x^2$. Then we define
\begin{equation}
  V_3 := \;\CA_3^\sharp\;,\quad V_4 :=\; \CA_4^\sharp\;,
\end{equation}
where $\sharp$ is the map from $T^*\CE_K$ to $T\CE_K$ induced by the symplectic structure \eqref{Esymplectic}. To get a pair of solutions to the section condition, the task is therefore to find projections $\hat \pr_1$ from $\CC^\infty(\CE_K)$ to a subalgebra determined by the kernel of the derivations $D_3$ and $V_3$, and similar $\hat \pr_2$ to a subalgebra determined by the kernel of the derivations $D_4$ and $V_4$. In our example it is easy to find two such projections, the generators of the subalgebras denoted by a tilde:
\begin{align}\label{projections}
  \hat\pr_1&(x^1,x^2,x^3,x^4,\xi^1,\xi^2,\zeta_1,\zeta_2,\theta_+^3,\theta_+^4,p_1,p_2,p_3,p_4)\nonumber \\
  &=(\tilde x^1,\tilde x^2, \tilde x^3,\tilde \xi^1,\tilde \xi^2,\tilde \xi^3,\tilde \zeta_1,\tilde \zeta_2, \tilde \zeta_3,\tilde p_1, \tilde p_2, \tilde p_3)\nonumber \\
  &=(x^1,x^2,x^3,\xi^1,\xi^2,\theta_+^3,\zeta_1, \zeta_2, \theta_+^4, p_1, p_2-kx^1p_4, p_3)\nonumber \\
  \hat\pr_2&(x^1,x^2,x^3,x^4,\xi^1,\xi^2,\zeta_1,\zeta_2,\theta_+^3,\theta_+^4,p_1,p_2,p_3,p_4) \nonumber \\
  &= (\tilde x^1,\tilde x^2, \tilde x^3,\tilde \xi^1,\tilde \xi^2,\tilde \xi^3,\tilde \zeta_1,\tilde \zeta_2, \tilde \zeta_3,\tilde p_1, \tilde p_2, \tilde p_3)\nonumber \\
  &=(x^1,x^2,x^4,\xi^1, \xi^2, \theta_+^4,\zeta_1,\zeta_2,\theta_+^3, p_1, p_2-jx^1 p_3, p_4)\;.
\end{align}
Using the periodicity \eqref{Eperiod} together with \eqref{finalcoord} we identify the image of the projections as the two Courant algebroids we started from, confirming that the procedure we suggest gives the right results at least in the easy example case. We have
\begin{equation}
\begin{aligned}
  &\hat \pr_1 : \CC^\infty(\CCE_K) \rightarrow \CC^\infty(\CCC_{j,k})\;, \\
  &\hat \pr_2 : \CC^\infty(\CCE_K) \rightarrow \CC^\infty(\CCC_{k,j})\;.
\end{aligned}
\end{equation}
Conversely, we also can embed $\CC^\infty(\CCC_{j,k})$ and $\CC^\infty(\CCC_{k,j})$ into $\CC^\infty(\CCE_K)$ such that the embedding maps are sections of the corresponding projection maps, i.e. we find embeddings
\begin{equation}
  e_i : \CC^\infty(\CCC_{j,k}) \hookrightarrow \CC^\infty(\CCE_K)\;, \quad \CC^\infty(\CCC_{k,j}) \hookrightarrow \CC^\infty(\CCE_K)\;,
\end{equation}
such that $\hat \pr_i \circ e_i = \textrm{id}$. The explicit form of the embedding maps in our example is as follows:
\begin{align}\label{embeddings}
  &e_1(\tilde x^1,\tilde x^2, \tilde x^3, \tilde \xi^1,\tilde \xi^2, \tilde \xi^3, \tilde \zeta_1, \tilde \zeta_2, \tilde \zeta_3, \tilde p_1, \tilde p_2 \tilde p_3)\nonumber \\
  &=(x^1,x^2,x^3,x^4,\xi^1,\xi^2,\zeta_1,\zeta_2,\theta_+^3,\theta_+^4,p_1, p_2, p_3, p_4)\nonumber \\
  &=(\tilde x^1, \tilde x^2, \tilde x^3, -k\tilde x^1\tilde x^2, \tilde \xi^1, \tilde \xi^2, \tilde \zeta_1, \tilde \zeta_2, \tilde \xi^3, \tilde \zeta_3, \tilde p_1, \tilde p_2, \tilde p_3, 0)\nonumber \\
  &e_2(\tilde x^1,\tilde x^2, \tilde x^3, \tilde \xi^1,\tilde \xi^2, \tilde \xi^3, \tilde \zeta_1, \tilde \zeta_2, \tilde \zeta_3, \tilde p_1, \tilde p_2 \tilde p_3)\nonumber \\
  &=(x^1,x^2,x^3,x^4,\xi^1,\xi^2,\zeta_1,\zeta_2,\theta_+^3,\theta_+^4,p_1, p_2, p_3, p_4)\nonumber \\
  &=(x^1, x^2, -j\tilde x^1\tilde x^2, \tilde x^3, \tilde \xi^1, \tilde \xi^2, \tilde \zeta_1, \tilde \zeta_2, \tilde \zeta_3, \tilde \xi^3, \tilde p_1, \tilde p_2, 0, \tilde p_3)\;.
\end{align}
The embeddings are well-defined as one readily checks \eqref{Eperiod}. Furthermore $e_i$ preserve the structure of the respective pre-NQ manifolds. The symplectic structures and homological functions of the Courant algebroids on the nilmanifolds are pull-backs of the corresponding objects on $\CCE_K$. Note that the homological functions arise as pullbacks of a pre-homological function as the Poisson square of the latter lies in the kernel of the pullbacks.

Finally, we are able to reformulate T-duality between the circle bundles $N_{j,k}$ and $N_{k,j}$ of diagram \eqref{eq:generic_diagram} in the language of NQ-manifolds. We first have to lift \eqref{dualitycond} to the NQ-language. First we split the pre-homological function \eqref{prehom} into an untwisted part $\CQ_0$ and the flux contributions:
\begin{equation}
  \CQ_{\CCE_K} =\,\CQ_0 + k\xi^1\xi^2\theta_+^3 + j\xi^1\xi^2\theta_+^4\;.
\end{equation}
$\CQ_0$ represents the ordinary de Rham differential on the correspondence space. The condition \eqref{dualitycond} thus takes the form
\begin{equation}\label{dualitycond2}
  \lbrace \CQ_0 , \CF \rbrace = \, -k\xi^1\xi^2\theta_+^3 + j\xi^1\xi^2\theta_+^4\;.
\end{equation}
We note that the product of the lifted connections on the two circle bundles $-\pr^* A \wedge \tilde \pr^*\tilde A$ after projection to $\CCE_K$ takes the form $\CF = -\theta^3_+\theta^4_+$. We are now able to phrase the statement of T-duality of circle bundles in the following form:
\begin{framed}
  \noindent {\bf T-duality of $N_{j,k}$ and $N_{k,j}$} \newline
  
  
  \noindent Let $\CF = -\theta_+^3\theta_+^4$ be the lift of the product of the connection forms to $\CCE_K$. If $\lbrace \CQ_{\CCE_K}, \CF\rbrace = 0$ on $\CCE_K$, then the homological functions are related by
  \begin{equation}
    \CQ_{\CCC_{k,j}} =\; e_2^*\circ \hat \pr_1 ^* \,\CQ_{\CCC_{j,k}}\;,
  \end{equation}
  with $e_2$ of \eqref{embeddings} and $\hat \pr_1$ in \eqref{projections}.
\end{framed}
To see this, we observe that $\lbrace \CQ_{\CCE_K},\CF\rbrace = 0$ is equivalent to the duality condition \eqref{dualitycond2}, i.e. that $\hat \pr_1^* H = k\xi^1\xi^2\theta^3_+$ and $\hat \pr_2^* \tilde H = j\xi^1\xi^2\theta^4_+$ represent cohomologous elements on $K$. Evaluating it for $\CF = -\theta_+^3\theta_+^4$, we get
\begin{equation}\label{intermediate}
  \theta_+^3(p_3 + k\xi^1\xi^2) = \,\theta_+^4(p_4  + j\xi^1\xi^2)\;.
  \end{equation}
Now the pullback of $\CQ_{\CCC_{j,k}}$ can be transformed:
\begin{align}
  \hat \pr_1 ^*\,\CQ_{\CCC_{j,k}} =\,& \xi^1 p_1 + \xi^2 p_2 +\theta_+^3 p_3 -\xi^2 x^1 (kp_4 + jp_3) \nonumber \\
  &\hspace{4pt} + k\,\xi^1\xi^2\theta_+^3\nonumber \\
  =\,&\xi^1 p_1 + \xi^2 p_2 + \theta_+^4 p_4 -\xi^2 x^1(kp_4 + jp_3) \nonumber \\
  &\hspace{4pt}+ j\,\xi^1 \xi^2\theta_+^4\nonumber \\
  =\,&\hat \pr_2^* \CQ_{\CCC_{k,j}}\;.
\end{align}
Hence, $(e_2^*\circ \hat \pr_1^*) \CQ_{\CCC_{j,k}} = \CQ_{\CCC_{k,j}}$.

We see that in case of a T-dual pair, we can transport the homological functions from the original Courant algebroid to the dual one. We used the T-duality condition \eqref{dualitycond} and the form of the homological function on the correspondence space. The investigation of more complicated examples in this language is the next step for future work.

Finally, we comment on how the $H$-flux transforms under the map $e_2^* \circ \hat \pr_1^*$. Using \eqref{projections} and \eqref{embeddings} we have
\begin{equation}\label{fluxtrafo}
  e_2^* \circ \hat \pr_1 ^* \, H = e_2^*(k\,\xi^1\xi^2\theta_+^3) = k\,\tilde \xi^1\tilde \xi^2 \tilde \zeta_3\;.
\end{equation}
The latter is an object with components $f^3_{12}$ and its value $k$ determines the non-closedness of the third basis one-form on the nilmanifold $N_{k,j}$. This is what is sometimes called $f$-flux, which we get from the $H$-flux on $N_{j,k}$ by applying the map $e_2^*\circ \hat \pr_1^*$.

\section{Open problems}

After the interpretation of local quantities appearing in DFT such as the section condition or the C-bracket, we approached global aspects of DFT on 2-step nilmanifolds in the form of circle bundles over 2-tori. Obviously an important next step is to consider more general $k$-torus bundles with abelian gerbe structures, where T-duality was studied e.g. in \cite{Bouwknegt:2003zg}. First, generalizations of the projections and embeddings \eqref{projections} and \eqref{embeddings} should be found. In all cases, a more geometric interpretation of the projections in terms of symplectic graded reduction would be desirable. Second, the map $e_2^* \circ \hat \pr_1 ^*$ appears to encode properties of T-duality. If it exists in the more general cases, its effect on the 3-form flux representing the Dixmier Douady class of the underlying gerbe has to be studied. In the easy case of \eqref{fluxtrafo} it maps the $H$-flux into the $f$-flux. It is to be expected that in case of its existence it will have more exotic types of fluxes such as the nongeometric $Q$-flux in its image. A study of the coexistence of several types of fluxes on nilmanifolds was done in \cite{Chatzistavrakidis:2013wra}. In these more exotic cases which arise for less restrictive isometry properties of the 3-form $H$, it is intriguing to find the link from formulations on doubled spaces \cite{Hull:2007jy} to the mathematical results of \cite{Bouwknegt:2004ap}. The use of noncommutative or nonassociative algebras in the latter on the one hand and ordinary geometry on doubled spaces on the former (by however mixing momentum and winding coordinates) needs to be understood more precisely.

Another open question is to write down actions of the form \eqref{DFTaction} using (pre-) NQ-manifolds. Differential forms and integration on such types of manifolds is a highly nontrivial subject which was not applied to DFT as far as the author knows. Integration on supermanifolds was studied with applications to supergravity actions in \cite{Voronov:1996wy,Witten:2012bg,Castellani:2014goa,Castellani:2015paa,Castellani:2016ibp}. Of course the role of supermanifolds there is to incorporate supersymmetry rather than Courant algebroid structures, but the techniques are closely related. To extend our formalism to include supersymmetric versions of gravity is an additional goal for the future. But already in the pure bosonic case, a reproduction of \eqref{DFTaction} in a geometric way (i.e. from a Riemann tensor defined on the corresponing NQ manifold) would help to understand more complicated examples of dynamics on extended spacetimes as they arise e.g. in compactifications of M-theory (exceptional field theories).

The case of compactification of eleven dimensional supergravity on higher dimensional tori leads to generalized tangent bundles of the form
\begin{equation}
  TM \oplus \wedge^2 T^*M \oplus \wedge^5 T^*M \oplus \Bigl(T^*M\otimes \wedge^7 T^*M\Bigr)\,,
\end{equation}
whose interpretation in terms of symplectic graded geometry is not known so far. First steps were achieved in \cite{Arvanitakis:2018cyo}, but the general case is an open problem. Studying derived brackets, algebraic conditions as \eqref{alg} might give a more systematic insight into dualities of such theories (U-duality). Extensions of spacetime which parameterize diffeomorphisms as well as gauge transformations of the corresponding fluxes in M-theory (exceptional field theories \cite{Berman:2010is,Berman:2011pe,Berman:2011cg,Berman:2011jh,Berman:2013eva,Ashmore:2015joa,Palmkvist:2015dea,Cederwall:2017fjm,Cederwall:2018aab,Cederwall:2018kqk}) in their generalized diffeomorphisms are an interesting playground to generalize the constructions reviewed in this article.

Finally, the derived bracket form of Courant- and C-brackets by Poisson brackets on symplectic graded ma\-ni\-folds suggests that they arise as classical limits of some quantized versions. As graded versions of the Moyal-Weyl product exist, it is intriguing to ask about the properties of deformed Courant algebroids and C-bracket structures whose star-commutator has to its lowest order the derived brackets which were described in this article. A naive attempt to interpret such deformations as originating from higher derivative corrections of string theoretic origin was given in \cite{Deser:2014wva,Deser:2017fko}. Independent of a link to string theory, it would be exciting to study deformations of derived brackets in a systematic way.

\providecommand{\othercit}{}
\providecommand{\jr}[1]{#1}
\providecommand{\etal}{~et~al.}

\bibliography{allbibtex}

\bibliographystyle{prop2015}

\end{document}